\documentclass[pra,preprint,showpacs]{revtex4}
\usepackage[centertags]{amsmath}
\usepackage{amsfonts}
\usepackage{amssymb}
\usepackage{amsthm} 
\usepackage{newlfont}
\usepackage{epsfig}
\usepackage{amscd}
\usepackage{graphicx}
\usepackage{epsfig}
\usepackage{footnote}
\usepackage{lipsum}
\usepackage{color}
\usepackage{xcolor}
\usepackage{graphicx}
\usepackage{multirow}
\usepackage{subfig}
\usepackage{amscd}

\newcommand{\beq}{\begin{equation}}
\newcommand{\eeq}{\end{equation}}
\newcommand{\ba}{\begin{array}}
\newcommand{\ea}{\end{array}}
\newcommand{\bea}{\begin{eqnarray}}
\newcommand{\eea}{\end{eqnarray}}
\begin{document}

\begin{center}
{\large \sc \bf {Partial structural  restoring of  two-qubit transferred state}
}

\vskip 15pt

%{\large 
%G.A.Bochkin, S.I.Doronin and 
A.I.Zenchuk 
%}

\vskip 8pt

{\it $^2$Institute of Problems of Chemical Physics RAS,
Chernogolovka, Moscow reg., 142432, Russia}

\end{center}
%\today

\begin{abstract}
We consider the communication line with two-qubit sender and receiver, the later is embedded into the four-qubit extended receiver. 
Using the  optimizing unitary transformation on the extended receiver  we  restore the structure of the non-diagonal part of an arbitrary 
initial sender's  state at the remote receiver at certain time instant. Obstacles for restoring the diagonal part are discussed.
We represent examples of such structural restoring in a  communication line of  42 spin-1/2 particles.
\end{abstract}

\maketitle

%%%%%%%%%%%%%%%%
\section{Introduction}
\label{Section:Introduction}

The problem of remote state creation originates from the problem of pure  state transfer formulated by Bose \cite{Bose} and becomes an attracting branch in the area of quantum information processing. Apparently,  the state initially created at the sender of a communication line can  not be transferred  to  the receiver unless  special protocols are  implemented.  
Among the first protocols we refer to that of  perfect state transfer \cite{CDEL,ACDE,KS}, remote boundary \cite{GKMT,WLKGGB} and optimized boundary \cite{BACVV2010,ZO,BACVV2011,ABCVV,SAOZ} state transfer.  Later the remote state creation protocols have been proposed and first realized for the photon systems  \cite{PBGWK2,PBGWK,XLYG}, where photons  are considered as a basic couriers of quantum information over  a long distance. However, the short distance information transfer in quantum information devices can be based on different objects, such as spin chains. As for remote creating  a one-qubit state in a spin system, the creatable region in the receiver's  state space  can be completely described  \cite{Z_2014,BZ_2015} because the one-qubit state-space is parametrized with only three parameters. 
Thus the one-to-one mapping 
\begin{eqnarray}\label{cr}
{\mbox{initial sender's state}}\;\;\to \;\; {\mbox{creatable receiver's state}}
\end{eqnarray}
is established in that case. 
The complete characterization of  creatable region in  higher dimensional  state-space is  much more complicated. Even two-qubit state depends on 15 parameters, so that  the mapping (\ref{cr})  can be hardly visualized. Although we  can construct certain families of states in this case, like Werner states \cite{Werner} in Ref.\cite{SZ_2016}, finding the protocol allowing a more careful control of the link between the initial sender's state and the creatable receiver's state 
is meaningful. 

A method of such control is proposed in Ref.\cite{FZ_2017,BFZ_Arch2018}
where the creation of a two-qubit block-scaled states is considered. In this case the receiver's state defers from the sender's one by the factor ahead of certain blocks of the density matrix. These blocks are multiple-quantum (MQ) coherence matrices. Remember that  the $n$-order coherence matrix collects  those elements of the density matrix which are responsible for the state-transitions  changing the $z$-projection of the total spin by $n$). The feature of these blocks is that they evolve independently provided that the 
dynamics is described by the Hamiltonian conserving the  $z$-projection of the total spin momentum. However, the protocol proposed in that paper requires a special initial state and is  not applicable to  an arbitrary one. As a result, each  MQ-coherence matrix caries at most one arbitrary parameter.
%except the zero-order coherence matrix which must be completely fixed initially in the sender's state.  

In this paper we modify the mentioned protocol { by} implementing the extended receiver (the subsystem at the receiver side embedding the receiver itself \cite{BZ_2016}) and the {  fixed} optimizing unitary transformation on it. {  We emphasize that, being fixed, the optimizing unitary transformation represents a part of the protocol and  remains the same for any transferred state.} As a  result we manage to structurally reconstruct the non-diagonal {  part} of the initial  sender's density matrix 
in the receiver's density matrix at certain time instant, i.e., the non-diagonal elements of the receiver's density matrix  become proportional to the appropriate elements of the  sender's initial density matrix in our protocol. We also discuss the obstacle arising in restoring the diagonal elements of the initial sender's density matrix. 

%{  In addition, our protocol updates the protocol of information extracting from the receiver state 
% formulated in Ref.\cite{Z_JPA_2012}. In that paper it is shown that all elements of the  sender's initial density matrix appear as  independent  linear %combinations in the elements of
% the receiver's density matrix. Thus the problem of the sender's density matrix transfer is reduced to the 
%linear  system of equations. The structural restoring in our protocol means that the above system of equations can be resolved for the non-diagonal part %of the density matrix. }

{  Let us mention another aspect of our protocol. Apparently,  if there is no optimizing unitary transformation, then the elements of the sender's initial density matrix appear in the receiver's density matrix as linear combinations. Thus, the problem of state transfer reduces to  the system of linear algebraic equations for 
the elements of the sender's density matrix which is solvable in general \cite{Z_JPA_2012}. However, the quantum-mechanical solution of this system in spin-1/2 communication line is not proposed.  Implementing the optimizing unitary transformation, we solve the nondiagonal part of this system and find scaled 
nondiagonal  elements of the sender's initial density matrix. Therefore our protocol contributes into the problem of solving the linear algebraic equations via quantum-mechanical methods \cite{HHL}. 
}

The paper is organized as follows.
The proposed model of a communication line is described in Sec.\ref{section:model}. The general protocol of structural restoring of a two-qubit state,
including the optimization of the   time instant for state registration and construction of  the optimal unitary transformation on the four-qubit  extended receiver, 
is proposed in Sec.\ref{section:restoring}. Examples of  numerical structural restoring of non-diagonal elements of a two-qubit initial sender's state  in the communication line of $N=42$ nodes  are represented in Sec.\ref{section:numerics}. 
General conclusions are given in Sec.\ref{section:conclusion}.

\section{Model}
\label{section:model}

\begin{figure*}
\epsfig{file=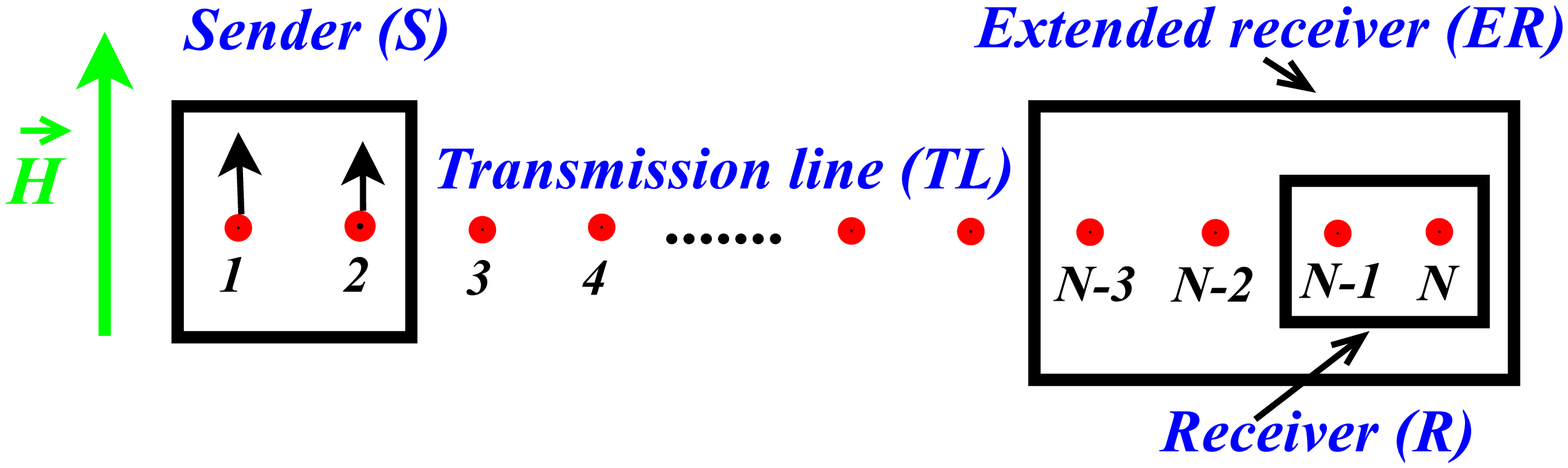,%communication2018,
  scale=0.5
   ,angle=0%270
}  
\caption{Communication line including the sender (S), transmission line (TL),  receiver (R) and extended receiver (ER). $\vec H$ is the external magnetic field.}
  \label{Fig:CL} 
\end{figure*}

We consider the communication line consisting of   two-qubit sender $S$  and  two-qubit receiver $R$ embedded into the four-node extended receiver $ER$, which is connected to the sender  through the transmission  line $TL$, see Fig.\ref{Fig:CL}.  
The spin dynamics is governed by the XX-Hamiltonian with the dipole-dipole interaction
\begin{eqnarray}\label{XY}
&&H=\sum_{j>i} D_{ij} (I_{ix}I_{jx} +I_{iy}I_{jy}),\\\label{comm}
&&[H,I_z]=0,
\end{eqnarray}
where $D_{ij}=\frac{\gamma^2 \hbar}{r_{ij}^3}$ is the coupling constant between 
the $i$th and $j$th nodes, $\gamma$ is the gyromagnetic ratio, $r_{ij}$ is the distance between the $i$th and $j$th nodes,
$I_{i\alpha}$ ($\alpha=x,y,z$)  is the projection operator of the $i$th spin on the $\alpha$ axis and $I_z=\sum_i I_{iz}$. We also consider the tensor-product initial state 
\begin{eqnarray}\label{in2}
\rho(0)=\rho^{(S)}(0) \otimes \rho^{(TL)} \otimes \rho^{(ER)}(0),
\end{eqnarray}
where $\rho^{(S)}(0)$ is an arbitrary  initial state of the sender $S$,  $\rho^{(TL)}$ and $\rho^{(ER)}$ are   the ground states (states without excitations) of the transmission line and extended receiver respectively:
\begin{eqnarray}
 \label{inTLB2}
\rho^{(TL)} &=&{\mbox{diag}}(1,0,\dots),\;\;\rho^{(ER)} ={\mbox{diag}}(1,0,\dots).
 \end{eqnarray}
Then
\begin{eqnarray}\label{rhoR}
\rho^{(R)}(t) = {\mbox{Tr}}_{rest} \Big(\tilde V(t)\rho(0)\tilde V^+(t)\Big),\;\;\ \tilde V(t)=e^{-i H t}.
\end{eqnarray}
 Here the trace is over all the nodes of the communication line except the receiver.

 In our protocol we use  the expansion of the density matrices in the sums of the multiple-quantum (MQ) coherence matrices \cite{FL} which read in the two-qubit case as
\begin{eqnarray}\label{MQ}
\rho^{(S)} = \sum_{k=-2}^2 \rho^{(S;k)},\;\;\;\rho^{(R)} = \sum_{k=-2}^2 \rho^{(R;k)},
\end{eqnarray}
where the MQ coherence matrices $\rho^{(S;k)}$ collect terms responsible for the state transfers   changing the $z$-projection of the  total spin momentum by $k$.

We emphasize that the commutation condition (\ref{comm}) together with the initial condition (\ref{inTLB2}) (where the initial density matrices $\rho^{(TL)}$ and $\rho^{(ER)}$ include only the zero-order coherence matrix) provides transferring
the MQ-coherence matrices from the sender to the receiver without mutual interaction \cite{FZ_2017}.

 \section{Structural restoring of transferred two-qubit  state }
 \label{section:restoring}
 
Hereafter we use the
Dirac notation for the basis in the state-space  of  $M$-qubit  (sub)system: 
\begin{eqnarray}\label{Dirac}
|n_1\dots n_M\rangle,\;\;n_i =  0, 1,\;\;i=1,\dots,M,
\end{eqnarray}
and the appropriate 
multi-index
$J_M=\{n_1\dots n_M\}$. For instance, the  elements of a density  matrix $\rho$ in the $M$-qubit state-space read
\begin{eqnarray}
\rho_{n_1\dots n_M;m_1\dots m_M} = \langle n_1\dots n_M|\rho | m_1\dots m_M\rangle.
\end{eqnarray}

We say that the initial sender's density matrix $\rho^{(S)}(0)$ is structurally restored  at the receiver if the  elements of the receiver's density matrix 
are proportional to the corresponding elements of the initial sender's density matrix up to the normalization condition \cite{FZ_2017}. For the two-qubit state restoring, by virtue of expansions (\ref{MQ}) for  the sender's and receiver's  density matrices, the restored state has the following form:
\begin{eqnarray}\label{restorred1}
 &&
\rho^{(R;k)}_{n_1n_2;m_1m_2}=
\lambda^{(k)}_{n_1n_2;m_1m_2} \rho^{(S;k)}_{n_1n_2;m_1m_2},\;\;\;k=\pm 1,\pm 2,\;\; m_1+m_2-n_1-n_2=k,\\\label{restorred2}
&&
\rho^{(R;0)}_{n_1n_2;m_1m_2}=\lambda^{(0)}_{n_1n_2;m_1m_2} \rho^{(S;0)}_{n_1n_2;m_1m_2},\;\;\;m_1+m_2-n_1-n_2=0,\\\label{norm}
&&
\rho^{(R;0)}_{00;00} = 1 - \sum_{k_1+k_2\neq 0} \lambda^{(0)}_{k_1k_2;k_1k_2} \rho^{(S;0)}_{k_1k_2;k_1k_2},\\\nonumber
&&
\lambda^{(-k)}_{m_1m_2;n_1n_2} = (\lambda^{(-k)}_{n_1n_2;m_1m_2})^*, \;\;k=0,1,2,\;\;
{\mbox{Im}}\lambda^{(0)}_{n_1n_2;n_1n_2} =0,
\end{eqnarray}
where normalization condition  (\ref{norm}) provides ${{\mbox Tr}}\rho^{(R)}=1$.
 
%%%%%%%%%%%%%%%%%%%%%%%
%%%%%%%%%%%%%%%%%%%%%%%
\subsection{Time instant for state registration}
\label{section:time}
\begin{figure*}
\epsfig{file=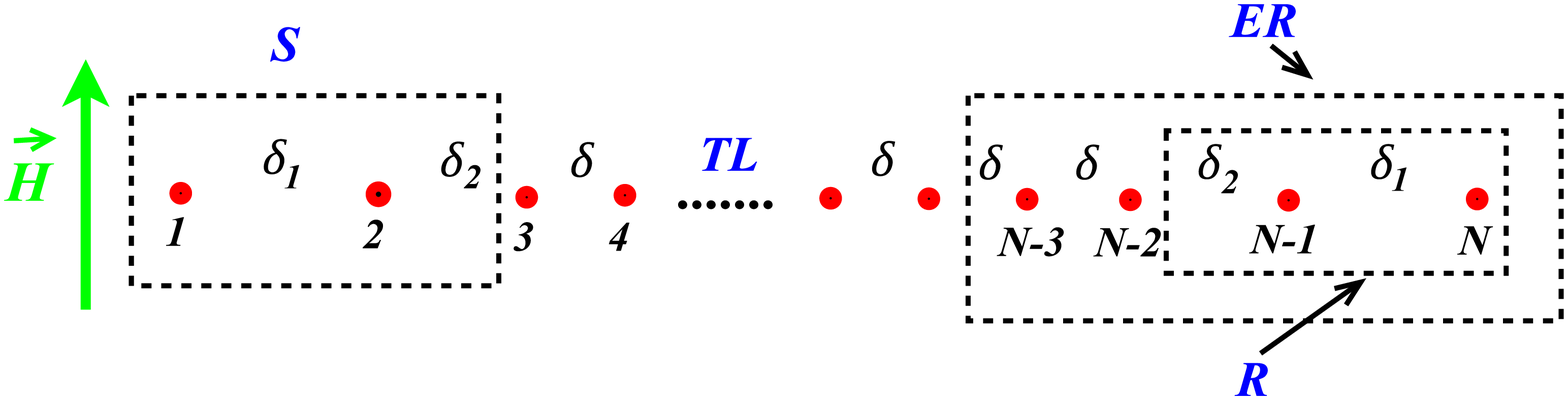, %communication2018opt,
  scale=0.5
   ,angle=0%270
}  
\caption{Communication line shown in Fig.\ref{Fig:CL} with optimized two pairs of  boundary coupling constants
$D_{1,2}=D_{N-1,N}=\delta_1$ and $D_{2,3}=D_{N-2,N-1}=\delta_2$.}
  \label{Fig:CC} 
\end{figure*}

To proceed with, we have to find the optimal time instant for  state registration at the receiver.
Since the second-order coherence intensity is, generally,  the smallest one in the two-qubit system, we select the   time instant corresponding to the maximal value of this quantity. Unlike the other coherences, the second-order coherence includes only one  element $ \rho^{(R)}_{00;11} \equiv \rho^{(R;2)}_{00;11}$ of the density matrix and therefore there is no mixing of elements in this matrix block. 
For this single element we have
\begin{eqnarray}
\label{seccoh}
&&
\rho^{(R)}_{00;11} = \langle 00 | \rho^{(R)}(t) |11\rangle = \\\nonumber
&&
\sum_{J_{N-2}}
\tilde V_{ J_{N-2} 00 ;0_{N}} \rho^{(S)}(0)_{00;11} \rho^{(TL)}_{0_{N-6};0_{N-6}}\rho^{(ER)}_{0_{4};0_{4}} \tilde V^+_{110_{N-2};  J_{N-2} 11}=
\rho^{(S)}_{00;11} \tilde V_{0_N; 0_N} \tilde V^+_{11 0_{N-2}; 0_{N-2} 11}.
\end{eqnarray}
Here  $0_K$ is the set of $K$ zeros.  
We also use the fact that  our system evolves  in  the two-excitation subspace due to commutation (\ref{comm}) and initial state (\ref{in2}), (\ref{inTLB2}).  Therefore, the sum over $J_{N-2}$ reduces to the single term with 
$J_{N-2}=0_{N-2}$. Since $\tilde V_{0_N; 0_N} =1$ (we set the 
energy of the  ground state to be zero, therefore this state does not evolve) we have  
\begin{eqnarray}
\rho^{(R)}_{00;11} = \rho^{(S)}_{00;11} \tilde V^+_{11 0_{N-2}; 0_{N-2} 11}. 
\end{eqnarray}
Consequently, to maximize the second-order coherence intensity  $I^{(2)}=|\rho^{(R)}_{00;11}|^2$ we have to maximize $|\tilde V_{11 0_{N-2}; 0_{N-2} 11}|^2$ which  is a smooth function of $t$ with the maximum 
\begin{eqnarray}
  %|\tilde V_{11 0_{N-2}; 0_{N-2} 11}|^2_{max}= 0.0150559 \;\;{\mbox{at}} \;\; t_{max}=46.0245
  |\tilde V_{11 0_{N-2}; 0_{N-2} 11}|^2_{max}= 0.0151 \;\;{\mbox{at}} \;\; t_{max}=46.0245
\end{eqnarray}
for the chain of $N=42$ nodes. To obtain the larger factor 
$|\tilde V_{11 0_{N-2}; 0_{N-2} 11}|^2_{max}$
we  optimize two boundary  pairs of the  coupling constants in the spin chain \cite{ABCVV}. 
For convenience, we introduce notation
\begin{eqnarray}
D_{i(i+1)}\equiv \delta_i,\;\;i=1,\dots,N-1.
\end{eqnarray}
For the homogeneous chain  we have $\delta_i=\delta$, $i=1,\dots,N-1$. For the chain with optimized boundary coupling constants we set
\begin{eqnarray}
 \delta_k=\delta,\;\; 3\le k\le N-3,\\\nonumber
 \delta_1 =\delta_{N-1},  \;\; \delta_2 =\delta_{N-2},
\end{eqnarray}
see Fig.\ref{Fig:CC}.
The maximization of $|\tilde V_{11 0_{N-2}; 0_{N-2} 11}|^2$ yields 
\begin{eqnarray}
%\delta_1 =0.3005 \delta, \;\; \delta_2 =0.5311\delta, \;\;  |\tilde V_{11 0_{N-2}; 0_{N-2} 11}|^2_{max}=0.437153, \;\;t_{max}=58.9826
\delta_1 =0.3005 \delta, \;\; \delta_2 =0.5311\delta, \;\;  |\tilde V_{11 0_{N-2}; 0_{N-2} 11}|^2_{max}=0.4372, \;\;t_{max}=58.9826
\end{eqnarray}
for $N=42$, i.e., $|\tilde V_{11 0_{N-2}; 0_{N-2} 11}|^2_{max}$ is increased by the factor of about 29 in comparison with the homogeneous chain. This optimized boundary chain is used below.
 
 %%%%%%%%%%%%%%%%%%%
 %%%%%%%%%%%%%%%%%%%
 \subsection{Optimizing unitary transformation}
 For restoring the structure of the  matrices $\rho^{(S;k)}$ at the receiver  we apply the optimizing  
local unitary transformation $V$ to the extended receiver, consisting of four nodes in our case. Thus, the total evolution operator $W$ reads
 \begin{eqnarray}\label{WV}
 W= V \tilde V,\;\;V=I_{2^{N-4}}\otimes V_0.
 \end{eqnarray}
 To avoid mixing the MQ-coherence matrices, the operator  $V_0$ has to preserve the $z$-projection of the total spin momentum, i.e.,
 \begin{eqnarray}\label{com}
 [V_0,I_z]=0.
 \end{eqnarray}
 To better characterize the structure of the unitary transformation, we pass to the scalar indexes from the Dirac notations (\ref{Dirac}) through the rule
 {\small
 \begin{eqnarray}\label{Dirac2}
\hspace*{-1cm}\begin{array}{|c|c|c|c|c|c|c|c|c|c|c|c|}
\hline
(n_1n_2n_3n_4)&(0000)&(0001)&(0010)&(0011)&(0100)&(0101)&(0110)&(1000)&(1001)&(1010)&(1100)\cr
i&1&2&3&4&5&6&7&8&9&10&11\cr
\hline
\end{array}
 \end{eqnarray}}
 and write the basis of the Lie algebra associated with this unitary transformation. This basis consists of 42 non-diagonal elements $\gamma^{(1;ij)}$, $\gamma^{(2;ij)}$, $j>i$ (diagonal elements are not useful in our transformations) with the following non-zero elements:
 \begin{eqnarray}
 \label{bases}
 &&
 \gamma^{(1;ij)}_{ij}= \gamma^{(1;ij)}_{ji}=1,\;\;\;\gamma^{(2;ij)}_{ij}= -\gamma^{(1;ij)}_{ji}=-i,\\\label{nonzerovarphi}
 (i,j)&\in&\{(2,3),(2,5),(2,8),(3,5), (3,8), (4,6), (4,7), (4,9), (4,10), (4,11), (5,8), \\\nonumber
 &&(6,7), (6,9), (6,10), (6,11), (7,9), (7,10), (7,11), (9,10), (9,11), (10,11)\}.
 \end{eqnarray}
 We represent the restoring operator $V_0$ in the form
 \begin{eqnarray}\label{V0}
 %V_0=e^{i \varphi^{(2;21)} \gamma^{(2;21)}} e^{i \varphi^{(1;21)}\gamma^{(1;21)}}\dots e^{i \varphi^{(2;1)} \gamma^{(2;1)}} e^{i \varphi^{(1;1)}
 %\gamma^{(1;1)}}
 V_0=\prod_{i=1}^{11}\prod_{j>i} e^{i \varphi^{(2)}_{i,j} \gamma^{(2;ij)}} e^{i \varphi^{(1)}_{i,j}\gamma^{(1;ij)}},
 % V_0&=&
 % e^{i \varphi^{(2;38)} \gamma^{(2;38)}} e^{i \varphi^{(1;38)}\gamma^{(1;38)}}
 % e^{i \varphi^{(2;35)} \gamma^{(2;35)}} e^{i \varphi^{(1;35)}\gamma^{(1;35)}}
 % e^{i \varphi^{(2;28)} \gamma^{(2;28)}} e^{i \varphi^{(1;28)}\gamma^{(1;28)}}    \times\\\nonumber
 %     &&
 % e^{i \varphi^{(2;25)} \gamma^{(2;25)}} e^{i \varphi^{(1;25)}\gamma^{(1;25)}}
 % e^{i \varphi^{(2;23)} \gamma^{(2;23)}} e^{i \varphi^{(1;23)}\gamma^{(1;23)}}
%,
 %INV: U5.U4.U3.U2.U1;
 %V_0&=&\prod_{{p,s=1}\atop{p\neq s}}^4 \exp(-i \varphi_{ps} I^+_p I^-_s)
 %\prod_{{p,s,r=1}\atop{p\neq s\neq r}}^4 \exp(-i \varphi_{psr} I^+_p I^-_s I_{zr})\times\\\nonumber
 %&&
 %\prod_{{p,s,r,q=1}\atop{p\neq s\neq r\neq q}}^4 \exp(-i \varphi_{psrq} I^+_p I^-_s I_{zr}I_{zq})
 %\prod_{{p,s,r,q=1}\atop{p\neq s\neq r\neq q}}^4 \exp(-i \tilde \varphi_{psrq} I^+_p I^-_s I^+_r I^-_q )
 \end{eqnarray}
 where  $\varphi^{(k)}_{i,j}$ are scalar parameters
and the product is  ordered in such a way that $i$ and $j$ increase from the right to the left. 

 \subsection{Restored   second-order coherence matrix}
 \label{section:restore2}
 
 The second-order coherence matrix doesn't require structural restoring because it consists of one element and therefore can be written in form (\ref{restorred1}):
 \begin{eqnarray}\label{2ord}
 &&
 \rho^{(R)}_{00;11} = W_{0_{N-2} 00 ;00 0_{N-2}} \rho^{(S)}_{00;11} W^+_{11 0_{N-2};0_{N-2}11}  =\lambda^{(2)}_{00;11} \rho^{(S)}_{00;11} ,\\\label{lamcoh2}
 &&
 \lambda^{(2)}_{00;11}=W^+_{11 0_{N-2};0_{N-2}11} .
  \end{eqnarray}
Here and below  we set $W_{0_{N-2} 00 ;00 0_{N-2}}=1$ because the ground state does not evolve. 
  
  %%%%%%%%%%%%%%%%%%%%%
  %%%%%%%%%%%%%%%%%%%%%
  \subsection{Structural restoring of  first-order coherence matrix}
  \label{section:restore1}

 The elements of the first-order coherence matrix after evolution and optimizing transformation  read:
 \begin{eqnarray}\label{1ord}
\rho^{(R)}_{00;n_1n_2} &=& \sum_{i_1+i_2=1}  \rho^{(S)}_{00;i_1i_2} W^+_{i_1i_2 0_{N-2};0_{N-2}n_1n_2}+ \\\nonumber
&&
\sum_{|J_{N-2}|=1}\sum_{i_1+i_2=1}  W_{J_{N-2}00 ;i_1i_2 0_{N-2}} \rho_{i_1i_2;11} W^+_{11 0_{N-2};J_{N-2}n_1n_2} , \\\label{1ord2}
\rho^{(R)}_{n_1n_2;11} &=& \sum_{i_1+i_2=1} W_{0_{N-2} n_1n_2 ;i_1i_2 0_{N-2}} \rho^{(S)}_{i_1i_2;11} W^+_{11 0_{N-2};0_{N-2}11},
 \end{eqnarray}
 where $|J_{N-2}|$ means the sum of the elements of the vector index  $J_{N-2}$ and $n_1+n_2=1$.
Deriving  these equations we take into account the relation
\begin{eqnarray}
W^+_{11 0_{N-2};J_{N-2}11}=W_{J_{N-2} 11 ;11 0_{N-2}} =0 \;\;{\mbox{if}} \;\;|J_{N-2}|\neq 0,
\end{eqnarray}
which is a consequence of  commutation relations (\ref{comm}) and (\ref{com}) (transitions changing the $z$-projection of the total spin momentum  are forbidden). 
 
The structural restoring of elements $\rho^{(S)}_{00;n_1n_2}$ and $\rho^{(S)}_{n_1n_2;11}$  (described by eqs.(\ref{1ord}) and 
(\ref{1ord2})) requires, respectively,
\begin{eqnarray}\label{1order}
&&\sum_{|J_{N-2}|=1} W_{J_{N-2}00 ;m_1m_2 0_{N-2}}  W^+_{11 0_{N-2};J_{N-2}n_1n_2}=0,\;\;n_1+n_2=m_1+m_2=1 \\\label{1order3}
\label{sol1ord1}
&&W^+_{n_1n_2 0_{N-2};0_{N-2}n_2n_1} =0 \;\; \Leftrightarrow \;\; W_{0_{N-2}n_2n_1;n_1n_2 0_{N-2}} =0,\;\;n_1+n_2=1.
\end{eqnarray}
Then equations (\ref{1ord}) and (\ref{1ord2}) get the structurally restored  form (\ref{restorred1}):
\begin{eqnarray}\label{1ordF}
\rho^{(R)}_{00;n_1n_2} &=& \lambda^{(1)}_{00;n_1n_2} \rho^{(S)}_{00;n_1n_2}, \\\nonumber
\rho^{(R)}_{n_1n_2;11} &=& \lambda^{(1)}_{n_1n_2;11} \rho^{(S)}_{n_1n_2;11} ,
 \end{eqnarray}
where $n_1+n_2=1$ and
\begin{eqnarray}\label{lamcoh1}
\lambda^{(1)}_{00;n_1n_2} = W^+_{n_1n_2 0_{N-2};0_{N-2}n_1n_2},\;\;\;\lambda^{(1)}_{n_1n_2;11} = W_{0_{N-2} n_1n_2 ;n_1n_2 0_{N-2}}W^+_{110_{N-2};0_{N-2}11}.
\end{eqnarray}
Thus, all in all, we have to solve  system (\ref{1order}), (\ref{sol1ord1}) of 6 complex equations for the $\varphi$-parameters of the unitary transformation $V_0$. This system is equivalent to the 12 real-valued equations which can be solved using at least  12 $\phi$-parameters of 42-parametric unitary transformation (\ref{V0}).  

%%%%%%%%%%%%%%%%%
%%%%%%%%%%%%%%%%%
\subsection{Structural restoring of  zero-order coherence matrix}
\label{section:restore0}
\label{section:zeroOrder}
The elements of the zero-order coherence matrix after the optimizing unitary transformation read:
\begin{eqnarray}
\label{0order2}
&&
\rho^{(R;0)}_{n_1n_2;m_1m_2}=\sum_{|J_{N-2}|=1}W_{J_{N-2}n_1n_2 ;11 0_{N-2}}\rho^{(S;0)}_{11;11} W^+_{11 0_{N-2}; J_{N-2} m_1m_2}+\\\nonumber
&&\sum_{{i_1+i_2=1}\atop{j_1+j_2=1}} 
W_{0_{N-2}n_1n_2 ;i_1i_2 0_{N-2}}\rho^{(S;0)}_{i_1i_2;j_1j_2} W^+_{j_1j_2 0_{N-2}; 0_{N-2} m_1m_2},\;\;n_1+n_2=m_1+m_2=1,
\\\label{0order3}
&&
\rho^{(R;0)}_{11;11}=\lambda^{(0)}_{11;11}\rho^{(S;0)}_{11;11} ,\;\;\lambda^{(0)}_{11;11} = W_{0_{N-2}11;11 0_{N-2}} W^+_{11 0_{N-2}; 0_{N-2} 11},\\
\label{0order1}
&&
\rho^{(R;0)}_{00;00}=\rho_{00;00}+\sum_{|J_{N-2}|=1} \sum_{{i_1+i_2=1}\atop{j_1+j_2=1}} W_{J_{N-2}00;i_1i_2 0_{N-2}} \rho^{(S;0)}_{i_1i_2;j_1j_2} W^+_{j_1j_20_{N-2};J_{N-2} 00}+
\\\nonumber
&&
\sum_{|J_{N-2}|=2} W_{J_{N-2}00;11 0_{N-2}} \rho^{(S;0)}_{11;11} W^+_{110_{N-2};J_{N-2} 00}.
 \end{eqnarray}
 These equations can be simplified in the case of structural restoring of the first order coherence matrix. In fact, 
 first, since  $W$ is a unitary transformation, we have
 \begin{eqnarray}\label{norm1}
 \sum_{|J_{N-2}|=1}  W_{J_{N};n_1n_2 0_{N-2}}  W^+_{n_2n_10_{N-2};J_{N}} = 0.
 \end{eqnarray}
Furthermore, in view of restoring conditions (\ref{sol1ord1}), eq.(\ref{norm1}) reduces to the form
 \begin{eqnarray}\label{norm2}
 %\sum_{|J_{N-2}|=1}  W_{J_{N-2} 00 ;i_1i_2 0_{N-2}}  W^+_{j_1j_20_{N-2};J_{N-2} 00}+
 %|W_{0_{N-2} i_1i_2 ;i_1i_2 0_{N-2}}|^2  % W^+_{i_1i_20_{N-2};0_{N-2}i_1i_2 } 
 %\delta_{i_1j_1}\delta_{i_2j_2}
 %= \delta_{i_1j_1} \delta_{i_2j_2}
 \sum_{|J_{N-2}|=1}  W_{J_{N-2} 00 ;n_1n_2 0_{N-2}}  W^+_{n_2n_10_{N-2};J_{N-2} 00}=0.
 \end{eqnarray}
 Finally, using eqs.(\ref{sol1ord1})  and (\ref{norm2}), respectively,  in eqs. (\ref{0order2}) and 
(\ref{0order1})  we write them as
 \begin{eqnarray}
 \label{0order22}
 &&
\rho^{(R;0)}_{n_1n_2;m_1m_2}=\sum_{|J_{N-2}|=1}W_{J_{N-2}n_1n_2 ;11 0_{N-2}}\rho^{(S;0)}_{11;11} W^+_{11 0_{N-2}; J_{N-2} m_1m_2}+\\\nonumber
&&
W_{0_{N-2}n_1n_2 ;n_1n_2 0_{N-2}} \rho^{(S;0)}_{n_1n_2;m_1m_2}  W^+_{m_1m_2 0_{N-2}; 0_{N-2} m_1m_2} ,\;\;n_1+n_2=m_1+m_2=1,\\\label{0order12}
&&
\rho^{(R;0)}_{00;00}=\rho^{(S;0)}_{00;00}+ \sum_{{i_1+i_2=1}} 
\sum_{|J_{N-2}|=1} |W_{J_{N-2}00;i_1i_2 0_{N-2}}|^2 \rho^{(S;0)}_{i_1i_2;i_1i_2} %W^+_{i_1i_2 0_{N-2};J_{N-2} 00}%+\right.
+\\\nonumber
&&
\sum_{|J_{N-2}|=2} |W_{J_{N-2}00;11 0_{N-2}}|^2 \rho^{(S;0)}_{11;11}.
 \end{eqnarray}
Eq.(\ref{0order12}) is  normalization trace-condition (\ref{norm}) and therefore does not require restoring. While restoring of  eq.(\ref{0order22})  implies the following constraints  for the operator $W$:
 \begin{eqnarray}\label{0constrW}
&&
\sum_{|J_{N-2}|=1}
W_{J_{N-2}n_1n_2 ;11 0_{N-2}} W^+_{11 0_{N-2}; J_{N-1}m_1m_2}=0,\;\; n_1+n_2=m_1+m_2=1. %\;\; n_1\neq m_1.
 \end{eqnarray}
System (\ref{0constrW}) combines two different  cases $(m_1,m_2)  = (n_1,n_2)$ and  $(m_1,m_2)  = (n_2,n_1)$. 
Writing  these two cases separately ($n_1+n_2=1$), 
\begin{eqnarray}\label{spl1}
&&
\sum_{|J_{N-2}|=1}
W_{J_{N-2}n_1n_2 ;11 0_{N-2}} W^+_{11 0_{N-2}; J_{N-1}n_2n_1}=0,\\\label{spl2}
&&
\sum_{|J_{N-2}|=1}
|W_{J_{N-2}n_1n_2 ;11 0_{N-2}} |^2=0,
\end{eqnarray}
we see that  eq.(\ref{spl2}) is a sum of positive terms. Therefore it is equivalent to the following system
\begin{eqnarray}\label{spl22}
W_{J_{N-2}n_1n_2 ;11 0_{N-2}}=0,\;\;n_1+n_2=1, \;\;|J_{N-2}|=1.
\end{eqnarray}
Moreover, if eqs.(\ref{spl22})  are satisfied, then eqs.(\ref{1order}) and (\ref{spl1}) become identities. 
All in all, the conditions for the complete structural restoring of both the first- and zero-order coherence matrices read ($n_1+n_2=1$)
\begin{eqnarray}\label{sol1ord12}
&&W_{0_{N-2}n_2n_1;n_1n_2 0_{N-2}} =0,\\\label{spl222}
&&W_{J_{N-2}n_1n_2 ;11 0_{N-2}}=0,\;\;|J_{N-2}|=1.
\end{eqnarray}
As a result, eqs. (\ref{0order12}) and (\ref{0order22}) get the structurally restored forms (\ref{restorred2}) and (\ref{norm}):
\begin{eqnarray}
\label{req2}
&&
%\rho^{(R;0)}_{n_1n_2;m_1m_2}=
%\lambda^{(0)}_{n_1n_2;m_1m_2} \rho^{(S;0)}_{n_1n_2;m_1m_2} ,\;\;n_1+n_2=1,\;\;m_1+m_2=1\\
\rho^{(R;0)}_{n_1n_2;n_1n_2}=
\lambda^{(0)}_{n_1n_2;n_1n_2} \rho^{(S;0)}_{n_1n_2;n_1n_2} ,\;\;n_1+n_2=1,\\
&&
\rho^{(R;0)}_{n_1n_2;n_2n_1}=
\lambda^{(0)}_{n_1n_2;n_2n_1} \rho^{(S;0)}_{n_1n_2;n_2n_1} ,\;\;n_1+n_2=1,\\
&&
\rho^{(R;0)}_{11;11}=\lambda^{(0)}_{11;11}\rho^{(S;0)}_{11;11},\\\label{0order12F}&&
\rho^{(R;0)}_{00;00}=\rho^{(S;0)}_{00;00}- \sum_{{i_1+i_2=1}} \lambda^{(0)}_{i_1 i_2;i_1i_2} \rho^{(S;0)}_{i_1i_2;i_1i_2}-
\lambda^{(0)}_{11;11} \rho^{(S;0)}_{11;11}
\end{eqnarray}
where 
 \begin{eqnarray}\label{lam01}
 &&
 \lambda^{(0)}_{n_1n_2;n_2n_1} = W_{0_{N-2}n_1n_2 ;n_1n_2 0_{N-2}}  W^+_{n_2n_1 0_{N-2}; 0_{N-2} n_2n_1}, \;\;
 n_1+n_2=1,\\\label{lam02}
 &&
 \lambda^{(0)}_{n_1 n_2;n_1n_2} = %-\sum_{|J_{N-2}|=1}  |W_{J_{N-2} 00;i_1i_2 0_{N-2}}|^2 = 
 |W_{0_{N-2}n_1n_2 ;n_1n_2 0_{N-2}}|^2 ,\;\;n_1+n_2=1,\\\label{lam03}
 &&
 \lambda^{(0)}_{11;11}=%-\sum_{|J_{N-2}|=2} |W_{J_{N-2}00;11 0_{N-2}}|^2 =
 |W_{0_{N-2}11;11 0_{N-2}}|^2 % W^+_{11 0_{N-2}; 0_{N-2} 11}.
\end{eqnarray}

\subsubsection{Obstacle for complete structural restoring}
\label{section:obst}
However, the direct analysis of the operator $W$  shows that  the following  subsystem of system (\ref{sol1ord12}), (\ref{spl222})
\begin{eqnarray}\label{I}
&&W_{0_{N-2}n_1n_2;n_2n_1 0_{N-2}} =0,\\\nonumber
&&W_{J_{N-4}00n_1n_2 ;11 0_{N-2}}=0, \;\;|J_{N-4}|=1,
\end{eqnarray}
where $n_1+n_2=1$, is not solvable in general. In fact,  
this subsystem consists of $2+2 (N-4)$ complex equations. According to (\ref{WV}), all  the elements 
of $W$ appearing in (\ref{I})  include the
elements $(V_0)_{00n_1n_2;i_1i_2i_3i_4}$ of the matrix $V_0$ which introduces $\varphi$-parameters. 
The direct analysis of  $V_0$  satisfying commutation condition (\ref{com}) shows that   
the above elements of $V_0$  depend on the following set of ten $\varphi$-parameters:
\begin{eqnarray}
\varphi^{(k)}_{2,3},
\varphi^{(k)}_{2,5},
\varphi^{(k)}_{2,8},
\varphi^{(k)}_{3,5},
\varphi^{(k)}_{3,8},\;\;k=1,2.
\end{eqnarray}
Therefore, system (\ref{I}) can not be solved even for the shortest chain of $N=6$ nodes (2-qubit sender and 4-qubit extended receiver), when the number of complex equations in (\ref{I}) 
is six.
Therefore,  solvability of (\ref{I}) requires increasing the dimensionality of the extended receiver.
In other words,  the dimensionality of the extended receiver required for the complete structural restoring of the diagonal part of the zero-order coherence matrix increases with the length $N$ of the communication line.  Such restoring  is not considered in our paper.

\subsubsection{Structural restoring of non-diagonal elements}
The obstacle pointed out in Sec.\ref{section:obst} doesn't arise in restoring the nondiagonal part of the zero-order coherence matrix. 
 Eqs.(\ref{sol1ord1}) and (\ref{spl1}) providing such restoring read ($n_1+n_2=1$):
\begin{eqnarray}\label{PartRestore0}\label{constrND1}
%&&\sum_{|J_{N-2}|=1} W_{J_{N-2}00 ;m_1m_2 0_{N-2}}  W^+_{11 0_{N-2};J_{N-2}n_1n_2}=0, \\\nonumber
&&W_{0_{N-2}n_1n_2;n_2n_1 0_{N-2}} =0,\\\label{constrND2}
&&
\sum_{|J_{N-2}|=1}
W_{J_{N-2}n_1n_2 ;11 0_{N-2}} W^+_{11 0_{N-2}; J_{N-1}n_2n_1}=0.
\end{eqnarray}
As the result, the 
 elements of the  partially restored  zero-order coherence matrix (Eqs. (\ref{0order22}), (\ref{0order3}) and (\ref{0order12})) read ($n_1+n_2=1$):
\begin{eqnarray}\label{req3}
&&
\rho^{(R;0)}_{n_1n_2;n_2n_1}=
\lambda^{(0)}_{n_1n_2;n_2n_1} \rho^{(S;0)}_{n_1n_2;n_2n_1} ,\\\label{req22}
&&
\rho^{(R;0)}_{n_1n_2;n_1n_2}= %\sum_{|J_{N-2}|=1} |W_{J_{N-2}n_1n_2 ;11 0_{N-2}}|^2 
\tilde\lambda^{(0)}_{n_1n_2;11}\rho^{(S;0)}_{11;11}+
  %|W_{0_{N-2}n_1n_2 ;n_1n_2 0_{N-2}}|^2
  \lambda^{(0)}_{n_1n_2;n_1n_2} \rho^{(S;0)}_{n_1n_2;n_1n_2},\\\label{req4}
&&
\rho^{(R;0)}_{11;11}=\lambda^{(0)}_{11;11}\rho^{(S;0)}_{11;11},\\\label{0order12F2}
&&
\rho^{(R;0)}_{00;00}=\rho^{(S;0)}_{00;00}- \sum_{{i_1+i_2=1}} \lambda^{(0)}_{i_1i_2;i_1i_2} \rho^{(S;0)}_{i_1i_2;i_1i_2},
-(\lambda^{(0)}_{11;11}+\tilde\lambda^{(0)}_{01;11}+\tilde\lambda^{(0)}_{10;11}) \rho^{(S;0)}_{11;11} ,
\end{eqnarray}
where $\lambda^{(0)}_{ij}$ are given in eqs.(\ref{lam01}) - (\ref{lam03}) and  
\begin{eqnarray}\label{tlam}
\tilde\lambda^{(0)}_{n_1n_2;11} =\sum_{|J_{N-2}|=1} |W_{J_{N-2}n_1n_2 ;11 0_{N-2}}|^2.
\end{eqnarray}

\subsection{Results on structural restoring of non-diagonal part of $\rho^{(S)}$}
 \label{section:results}
 Now we collect the results of Secs.\ref{section:restore2}-\ref{section:restore0} on structural restoring of the non-diagonal part of the initial sender's density matrix.
 
 The constraints for the elements of $W$, which must be resolved for $\varphi$-parameters of the unitary transformation (\ref{V0}) 
 and which provide the required restoring, are given by  Eqs.(\ref{1order}), (\ref{sol1ord1}) (or (\ref{constrND1})) and (\ref{constrND2}). For convenience, we collect them:
\begin{eqnarray}\label{PartRestore}
&&\sum_{|J_{N-2}|=1} W_{J_{N-2}00 ;m_1m_2 0_{N-2}}  W^+_{11 0_{N-2};J_{N-2}n_1n_2}=0,\;\;m_1+m_2=n_1+n_2=1, \\\nonumber
&&W_{0_{N-2}n_1n_2;n_2n_1 0_{N-2}} =0,\;\;n_1+n_2=1,\\\nonumber
&&
\sum_{|J_{N-2}|=1}
W_{J_{N-2}n_1n_2 ;11 0_{N-2}} W^+_{11 0_{N-2}; J_{N-2}n_2n_1}=0,\;\;n_1+n_2=1.
\end{eqnarray}
This system consists of 7 complex equations for 42 $\varphi$-parameters in the unitary transformation (\ref{V0}) and therefore can be satisfied, which is confirmed below in Sec.\ref{section:numerics}.  We remark, that the number of the $\varphi$-parameters in the similar unitary transformation (\ref{V0}) associated with the three-qubit extended receiver is 12, which is not enough to solve 7 complex equations 
(\ref{PartRestore}). This fact justifies our choice of the unitary transformation on the four-qubit extended receiver as a minimal state-restoring tool.
 The restored elements are given by eqs.(\ref{2ord}), (\ref{1ordF}), (\ref{req3}) - (\ref{0order12F2}),
 where $\lambda^{(k)}_{n_1n_2;m_1m_2}$ are given in eqs.(\ref{lamcoh2}), (\ref{lamcoh1}) and (\ref{lam01})-(\ref{lam03}), and $\tilde\lambda^{(0)}_{n_1n_2;11}$ are defined in (\ref{tlam}). 
We notice that the number of equations in system (\ref{PartRestore}) representing the system of constraint  on $W$   does not depend on the  length $N$ of a communication line.
Therefore, if we disregard the structural restoring of the diagonal part of the zero-order coherence matrix, then the required dimensionality of the extended receiver does not depend on the length $N$ of the communication line as well.

 \section{Numerical partial structural restoring
 %in communication line of $N=42$ nodes
 }
 \label{section:numerics}
Now  we numerically construct the partially structurally restored states discussed in Sec.\ref{section:results} using the communication line of $N=42$ nodes.
We show that this can be performed using the  $\varphi$-parameters in the optimizing unitary transformation.  In addition, we would like to maximize the scale factors corresponding to the  restored elements, i.e., the factors$|\lambda^{(2)}_{00;11}|$, $|\lambda^{(1)}_{n_1n_2;m_1m_2}|$ and 
 $|\lambda^{(0)}_{{ 01;10}}|$ {  which can be considered as damping factors because all of  them are less than one by absolute value. The reason of such damping is dispersion of a propagating state. Of course, we would prefer the minimal damping (i.e. all scale factors approach one by absolute value). However,} it is natural that all of them can not reach their maximal values simultaneously. Therefore, we 
 find such $\varphi$-parameters that maximize the absolute value of a selected scale factor. The solution of this problem is not unique, therefore we perform 
 1000 numerical experiments and choose the case  corresponding to the maximal  sum of  absolute values of all other scale factors.  {  This requirement fixes the optimizing unitary transformation.}
 The results of such optimization are collected in Table \ref{Table:opt}, where each of the  1st to 6th  rows  includes the scale factors 
 corresponding to   the maximization of, respectively, $|\lambda^{(0)}_{01;10}|$, $|\lambda^{(1)}_{00;01}|$, $|\lambda^{(1)}_{00;10}|$, 
 $|\lambda^{(1)}_{01;11}|$, $|\lambda^{(1)}_{10;11}|$ and $|\lambda^{(2)}_{00;11}|$. The 7th row in this table corresponds to the maximization of the sum of all above scale factors. Again, we perform 1000 numerical experiments and choose the case  corresponding to the maximal value of the minimal scale factor. We see that the  scale factors shown 
 in the 7th row are very close to those shown in the first row corresponding to the optimization of the scale factor $\lambda^{(0)}_{01;10}$. Moreover,  the absolute values of all scale factors are valuable {  and the most uniform} in these two rows.

{  As an example we represent the  set of $\varphi$-parameters for the optimizing unitary transformation (\ref{V0}) corresponding to the 7th row of  Table \ref{Table:opt}:
\begin{eqnarray}
\label{opt}
\begin{array}{cc}
\begin{array}{ll}
\varphi^{(1)}_{2,3}= 3.2173, & \varphi^{(2)}_{2,3} = 4.6606, \cr
 \varphi^{(1)}_{2,5} = 1.5820,& \varphi^{(2)}_{2,5} = 4.1773, \cr
 \varphi^{(1)}_{2,8} = 1.9863, &\varphi^{(2)}_{2,8}= 3.8653 , \cr
 \varphi^{(1)}_{3,5} = 2.9836, &\varphi^{(2)}_{3,5} = 2.8152, \cr
 \varphi^{(1)}_{3,8} = 1.6892,& \varphi^{(2)}_{3,8}= 1.5472, \cr
 \varphi^{(1)}_{4,6} = -0.0758,& \varphi^{(2)}_{4,6}= 5.9730, \cr
 \varphi^{(1)}_{4,7} = 2.9802, &\varphi^{(2)}_{4,7} = 6.5792, \cr
 \varphi^{(1)}_{4,9} = 3.3090, &\varphi^{(2)}_{4,9} = 3.4037, \cr
 \varphi^{(1)}_{4,10} = 1.9777,& \varphi^{(2)}_{4,10} = 2.0048, \cr
 \varphi^{(1)}_{4,11} = 1.5586,& \varphi^{(2)}_{4,11} = 4.5361 , \cr
 \varphi^{(1)}_{5,8} = 2.5019 ,& \varphi^{(2)}_{5,8} = 2.9224 , 
 \end{array} &
 \begin{array}{ll}
 \varphi^{(1)}_{6,7} = 3.1337, &\varphi^{(2)}_{6,7} = 3.2409 , \cr
 \varphi^{(1)}_{6,9} = 0.4114 ,& \varphi^{(2)}_{6,9} = 5.7006 , \cr
 \varphi^{(1)}_{6,10} = 1.7915 ,& \varphi^{(2)}_{6,10} = 4.5986 , \cr
 \varphi^{(1)}_{6,11} = 2.8966, &\varphi^{(2)}_{6,11} = 0.1255, \cr
 \varphi^{(1)}_{7,9} = 5.0837 ,& \varphi^{(2)}_{7,9} = 2.4180, \cr
 \varphi^{(1)}_{7,10} = 0.6389,& \varphi^{(2)}_{7,10} = 5.6101, \cr
 \varphi^{(1)}_{7,11} = 3.7066, &\varphi^{(2)}_{7,11} = 3.2203, \cr
 \varphi^{(1)}_{9,10} = 2.7775 ,& \varphi^{(2)}_{9,10} = 5.5717 , \cr
 \varphi^{(1)}_{9,11} = 0.5211, &\varphi^{(2)}_{9,11} = 6.2622 , \cr
 \varphi^{(1)}_{10,11} = 6.3862 , &\varphi^{(2)}_{10,11} = 4.2717. \cr
 ~~~
\end{array}
\end{array}
%SS1.dat
\end{eqnarray}
Implementing  this optimizing transformation  we perform mapping of an  arbitrary  nondiagonal elements of the density matrix $\rho^{(S)}(0)$ into the appropriate elements of the receiver's density matrix $\rho^{(R)}$ with the corresponding damping factors collected in the  7th row.  
}

 \begin{table}
 \begin{tabular}{|cccccc|}
 \hline
 $\lambda^{(0)}_{01;10}$ & $\lambda^{(1)}_{00;01}$   & $\lambda^{(1)}_{00;10}$  & $\lambda^{(1)}_{01;11}$  & $\lambda^{(1)}_{10;11}$&$\lambda^{(2)}_{00;11}$\cr
  \hline
$\mathbf{0.3501}e^{2.6200i}$ &$0.3871e^{0.5993i}$ & $0.9046 e^{-3.0639i}$ & $0.2138e^{-1.7284i}$ & $0.4996e^{1.9348i}$ & $0.5522e^{-1.1291i}$ \cr
$0.0044e^{-2.6047i}$ & $\mathbf{0.8122}e^{0.2813i}$ & $0.0055e^{-2.3234i}$ & $0.0948e^{2.3781i}$ & $0.0006e^{-1.3004i}$ & $0.1167e^{2.6594i}$ \cr
$0.0780e^{2.0820i}$ & $0.0833e^{1.1028i}$ & $\mathbf{0.9359}e^{-3.0983i}$ & $0.0066e^{-1.0999i}$ & $0.0744e^{3.1013i}$ & $0.0794e^{ 0.0029i}$ \cr
$0.0130e^{1.2029i}$ & $0.7883e^{-2.7766i}$ & $0.0165e^{-1.5737i}$ & $\mathbf{0.5525}e^{-0.1457i}$ & $0.0116e^{-1.3486i}$ & $0.7010e^{-2.9223i}$ \cr
$0.0557e^{ 3.0015i}$ & $0.0622e^{0.0781i}$ & $0.8953e^{3.0796i}$ & $0.0387e^{1.5415i}$ & $\mathbf{0.5568}e^{-1.4599i}$ & $0.6219e^{1.6197i}$ \cr
$0.0092e^{-1.7244i}$ & $0.2950e^{2.5666i}$ & $0.0312e^{0.8422i}$ & $0.2136e^{-0.7733i}$ & $0.0226e^{0.9510i}$ & $\mathbf{0.7239}e^{1.7932i}$ \cr
$0.3489e^{-1.7503i}$ & $0.3868e^{0.1608i}$ & $0.9019e^{-1.5895i}$ & $0.2201e^{0.2186i}$ & $0.5132e^{1.9689i}$ & $0.5690e^{0.3794i}$ \cr
\cr
\hline
 \end{tabular}
 \caption{\label{Table:opt} Optimized scale factors for the partial structural restoring of the two-qubit state in the communication line of $N=42$ nodes. The rows from 1 to 7  correspond  to the maximization of, respectively, $|\lambda^{(0)}_{01;10}|$, $|\lambda^{(1)}_{00;01}|$, $|\lambda^{(1)}_{00;10}|$, 
 $|\lambda^{(1)}_{01;11}|$, $|\lambda^{(1)}_{10;11}|$, $|\lambda^{(2)}_{00;11}|$  and the sum of the absolute values of all scale factors (the amplitudes of the  maximized scale factors  are in bold). In each case (except the last one)  we also maximize the sum of the absolute values of all other scale factors.
 }
 \end{table}
We notice also that the maximization of  $|\lambda^{(1)}_{00;01}|$, $|\lambda^{(1)}_{00;10}|$ and  $|\lambda^{(2)}_{00;11}|$ (see respectively the 2nd, 3rd and 6st rows in Table \ref{Table:opt}) makes these  factors  largest of all others. In other cases   the maximized scale factor is not the largest one, see the 1st, 4th and 5th rows in this table.

\section{Conclusions}
\label{section:conclusion}

A variant of remote state creation is the scaled state creation which is a simple well described  map of the initial sender's state into the receiver's one. In the ideal case this map reduces the state creation to the multiplication of each matrix element (except the  diagonal element providing normalization)   by a scalar scale factor and therefore can be regarded as a consequent development of the ideal state transfer protocol.  We refer to such map as a structural restoring of the initial sender's state. 

In our case of the partial structural restoring {  of the two-qubit density matrix}  the above scaling is implemented only for the non-diagonal elements, while   the diagonal elements remain uncontrolled. {  In addition, these scales are less then one by absolute value due to 
dispersion of a propagating state. This prompts us to call them the damping factors.}    In the proposed protocol  we  take into account the independent evolution of the MQ-coherence matrices.   The basic restoring tool is the unitary transformation of the so-called extended receiver consisting of four last nodes of the communication line and embedding the receiver itself. We show that the facility of  restoring  the non-diagonal part is independent on the total length of the communication line, unlike the diagonal part, whose restoring requires the larger  number of optimizing parameters in the unitary transformation  and this number  increases with $N$. The diagonal part restoring  is postponed for   further  study.  

We remark that the protocol of the block-scaled state transfer proposed in  Ref. \cite{BFZ_Arch2018} doesn't involve  the 
optimizing unitary transformation of the extended receiver. Those states have the same scale factor $\lambda^{(k)}$ for all the elements inside of each $\pm k$-order coherence matrix (up to the normalization of the zero-order coherence matrix). 
In addition,  each block $\rho^{(S;-1)}  + \rho^{(S;1)}$ and $\rho^{(S;-2)}  + \rho^{(S;2)}$  carry  only one arbitrary  scalar parameter, while the parameter of the zero-order coherence matrix is fixed by the requirement of maximizing the state-subspace covered by the parameters of the nonzero-order coherence matrices. 
In our paper,  the scale factors for all elements are  independent from each other and the first- and second-order coherence matrices carry, respectively, four and one complex parameters (remember that the zero-order coherence matrix is uncontrolled in our case). This enhances the capability  of our protocol as an  information transfer tool, while
additional relations among the scale factors   might  cause  appropriate relations among the  elements of $\rho^{(S)}(0)$ (which are completely independent in our case) and thus reduce the encoded information.  Nevertheless, studying  possible relations among the scale factors is a meaningful problem which deserves the further study.   

{ 
Finally, we notice that the optimizing unitary transformation used in the restoring protocol can be considered as a tool for solving the linear algebraic systems because it reduces the linear combinations of the elements of $\rho^{(S)}$ to the appropriate  elements multiplied by the damping  factors. The problem of reducing this damping appears to be of a principal meaning.}

This work is partially supported by the program of the Presidium of RAS No. 5 ''Electron resonance, spin-dependent electron effects and spin technology'' and by the Russian Foundation for Basic Research (Grant No.16-03-00056).


\begin{thebibliography}{99}
 
 
\bibitem{Bose}
S. Bose, Phys. Rev. Lett. {\bf   91}, 207901 (2003)

\bibitem{CDEL}
 M.Christandl, N.Datta, A.Ekert, and A.J.Landahl, Phys.Rev.Lett. {\bf   92}, 187902 (2004)

\bibitem{ACDE}
 C.Albanese, M.Christandl, N.Datta, and A.Ekert, Phys.Rev.Lett. {\bf   93}, 230502 (2004)

\bibitem{KS}
 P.Karbach and J.Stolze, Phys.Rev.A {\bf   72}, 030301(R) (2005)

 
\bibitem{GKMT}
 G.Gualdi, V.Kostak, I.Marzoli, and P.Tombesi, Phys.Rev. A {\bf   78}, 022325 (2008)
 
\bibitem{WLKGGB}
A.W\'ojcik, T.Luczak, P.Kurzy\'nski, A.Grudka, T.Gdala, and M.Bednarska
Phys. Rev. A {\bf   72}, 034303 (2005)

 \bibitem{BACVV2010}
 L. Banchi, T. J. G. Apollaro,  A. Cuccoli,  R. Vaia, and P. Verrucchi, 
 Phys.Rev.A {\bf 82}, 052321 (2010)
% Optimal dynamics for quantum-state and entanglement transfer
%through homogeneous quantum systems


\bibitem{ZO}
A. Zwick and O. Osenda, J. Phys. A: Math. Theor. {\bf 44}, (2011) 105302.


 \bibitem{BACVV2011}
 L. Banchi, T. J. G. Apollaro, A. Cuccoli, R. Vaia
and P. Verrucchi,
 New J. Phys. {\bf 13}, 123006 (2011) 
 %Long quantum channels for high-quality entanglement transfer
 % Entenglement transfer through the homogeneous spin chain with optimized boundary coupling

 
\bibitem{ABCVV}
T. J. G. Apollaro, L. Banchi, A. Cuccoli, R. Vaia, and
P. Verrucchi, Phys. Rev. A {\bf 85} (2012), 052319 
 
 
 \bibitem{SAOZ}
 J.Stolze, G. A. \'Alvarez,
O. Osenda, and  A. Zwick in
{\it Quantum State Transfer and Network Engineering.
Quantum Science and Technology},
ed. by  G.M.Nikolopoulos and I.Jex, Springer Berlin Heidelberg, Berlin, p.149  (2014) 

 
\bibitem{PBGWK2}
N.A.Peters, J.T.Barreiro,  M.E.Goggin, T.-C.Wei,  and P.G.Kwiat, Phys.Rev.Lett. {\bf   94}, 
150502 (2005) 
%Remote State Preparation: Arbitrary Remote Control of Photon Polarization

\bibitem{PBGWK}
N.A.Peters, J.T.Barreiro, M.E.Goggin, T.-C.Wei, and P.G.Kwiat in {\it Quantum
Communications and Quantum Imaging III}, ed. R.E.Meyers,
Ya.Shih, Proc. of SPIE {\bf   5893} (SPIE, Bellingham, WA, 2005) 
%doi:10.1117/12.615734
%Remote State Preparation: Arbitrary remote control of photon
%polarizations for quantum communication��. 

\bibitem{XLYG}
G.Y. Xiang, J.Li, B.Yu, and G.C.Guo
Phys. Rev. A {\bf   72}, 012315  (2005)
%Remote preparation of mixed states via noisy entanglement

 
\bibitem{Z_2014}
A.I.Zenchuk, 
%''Remote  creation of a one-qubit mixed state through a short homogeneous spin-1/2 chain'',
Phys. Rev. A {\bf 90}, 052302(13) (2014) 




\bibitem{BZ_2015}
G. A. Bochkin and A. I. Zenchuk, 
%Remote one-qubit-state control using the pure initial state of a two-qubit sender:
%Selective-region and eigenvalue creation,
Phys.Rev.A 91, 062326(11) (2015)

\bibitem{Werner}
R. F. Werner, Phys. Rev. A {\bf 40}, 4277 (1989)
%Quantum states with Einstein?Podolsky?Rosen correlations admitting a hidden-variable


 \bibitem{SZ_2016}
 J.Stolze and A.I.Zenchuk, Quant. Inf. Proc.  {\bf 15}, (2016) 3347
 %pp. 3347-3366 (2016)
 %Remote two-qubit state creation and its robustness, 
%DOI 10.1007/s11128-016-1345-5

 
\bibitem{FZ_2017}
E.B.Fel'dman, A.I.Zenchuk, JETP {\bf 125}, 1042 (2017)

 \bibitem{BFZ_Arch2018}
G.A.Bochkin, E.B.Fel'dman, A.I.Zenchuk, Quant.Inf.Proc.  {\bf 17}, 218 (2018)
%Transfer of scaled multiple-quantum coherence matrices

 \bibitem{BZ_2016}
 G. Bochkin and  A. Zenchuk, 
 %Extension of the remotely creatable  region via the local unitary
%transformation on the receiver side, 
Qunt. Inf. Comp. {\bf 16}, 1349 (2016) 
%1349-1364
%No. 15&16

{ 
\bibitem{Z_JPA_2012}
A.I.Zenchuk, J. Phys. A: Math. Theor. {\bf 45} 115306 (2012) 
%(11pp)
%Information propagation in a quantum system: examples of open spin-1/2 chains

\bibitem{HHL}
A. W. Harrow,  A. Hassidim,  and S. Lloyd, Phys.Rev.Lett. {\bf 103}, 150502 (2009)
%Quantum Algorithm for Linear Systems of Equations
}

 \bibitem{FL}
 E. B. Fel'dman, S. Lacelle, Chem. Phys. Lett. {\bf 253}, 27
(1996)


\end{thebibliography}
\end{document}